# Optimal Filtering for DDoS Attacks

Karim El Defrawy
ICS Dept.
UC Irvine
keldefra@uci.edu

Athina Markopoulou
EECS Dept.
UC Irvine
athina@uci.edu

Katerina Argyraki
EE Dept.
Stanford Univ.
argyraki@stanford.edu

## ABSTRACT
Distributed Denial-of-Service (DDoS) attacks are a major problem in the Internet today. In one form of a DDoS attack, a large number of compromised hosts send unwanted traffic to the victim, thus exhausting the victim's resources and preventing it from serving its legitimate clients. One of the main mechanisms that have been proposed to deal with DDoS is filtering, which allows routers to selectively block unwanted traffic. Given the magnitude of DDoS attacks and the high cost of filters in the routers today, the successful mitigation of a DDoS attack using filtering crucially depends on the efficient allocation of filtering resources. In this paper, we consider a single router, typically the victim's gateway, with a limited number of available filters. We study how to optimally allocate filters to attack sources, or entire domains of attack sources, so as to maximize the amount of good traffic preserved, under a constraint on the number of filters. We formulate the problem as an optimization problem and solve it optimally using dynamic programming, study the properties of the optimal allocation, experiment with a simple heuristic and evaluate our solutions for a range of realistic attack-scenarios. First, we look at a single-tier where the collateral damage is high due to the filtering at the granularity of domains. Second, we look at the two-tier problem where we have an additional constraint on the number of filters and the filtering is performed on the granularity of attackers and domains.

## 1. INTRODUCTION
Distributed Denial-of-Service attacks (DoS) are one of the most severe and hard to solve problems on the Internet today. During a DDoS attack, a large number of compromised hosts coordinate and send unwanted traffic to the victim thus exhausting the victim's resources and preventing it from serving its legitimate clients. For example, victims of DDoS attacks can be companies that rely on the Internet for their business, in which case DDoS attacks can result in severe financial loss or even in the company quitting the business [1]. Government sites (e.g. www1.whitehouse.gov) and other organizations can also be victims of DDoS attacks, in which case disruption of operation results in a political or reputation cost.

Several approaches and mechanisms have been proposed to deal with DDoS attacks; this is still an active research area with ongoing work. One body of anti-DDoS work has focused on developing DDoS detection mechanisms: how to quickly identify that an attack is ongoing, how to distinguish the legitimate from the attack traffic, and how to identify the paths where attack traffic is coming from. Another body of work focuses on DDoS defense mechanisms to mitigate the damage inflicted by a DDoS attack; defense mechanisms can be proactive, such as capabilities [11] [10] and/or re-active, such as filtering at the routers [8].

In this work, we are particularly interested in filtering mechanisms, which are a necessary component in the anti-DDoS solution. We consider the scenario of a bandwidth flooding attack, during which the bottleneck link to the victim is flooded with undesired traffic. To defend against such an attack, the victim must identify undesired traffic (using some identification mechanism which is not the focus of this work) and request from its ISP/gateway to block it before it enters the victim's access link and causes damage to legitimate traffic. Even assuming a perfect mechanism for identification of attack traffic, filter allocation at the victim's gateway is in itself a hard problem. The reason is that the number of attack sources in today's DDoS attacks is much larger than the number of expensive filters (ACLs) at the routers. Therefore, the victim cannot afford to selectively block traffic from each individual attack source, but instead may have to block entire domains; in that case legitimate traffic originating from that domain is also unnecessarily filtered together with the attack sources. Clearly, the successful mitigation of a DDoS attack using filtering, crucially depends on the allocation of filtering resources. In this paper, we do not propose a new architectural solution to DDoS. We study and optimize the use of an existing mechanism – filtering. Filters can be placed at a single gateways' tier, so as to maximize the preserved good traffic; the core insight in the single-tier problem is that the coarse filtering granularity makes co-located attack and legitimate traffic to share fate. This single-tier problem turned out to be a knapsack problem, with a greedy solution [21]. The insight in the multi-tier problem is between the preserved goodput and the number of filters used. In addition, we propose a simple two-tier heuristic and we evaluate our solution for three realistic attack scenarios, based on data sets from the analysis of the Code Red [16] and Slammer [17] worms, the Prolexic Zombie Report [19], and statistics on Internet users [20].

The structure of the rest of the paper is as follows. In section 2, we give more background on the problem and we discuss related work. In section 3, we formulate the problem of optimal filter allocation on a single tier (i.e. gateways or attackers tier) and the more general problem of filtering at both the gateway and attacker tier. We solve the problem optimally using dynamic programming. We study the properties of the optimal solution and develop a low complexity heuristic and compare its performance to that of the optimal solution through simulation in section 4. In section 5, we conclude the paper and discuss open issues and future work.

## 2. BACKGROUND
### 2.1 The DDoS Problem
There are several ways to launch DDoS attacks, which can be mainly classified into the following types. First, there are vulnerability attacks, when some vulnerability in the OS of the targeted machine or in the network stack is exploited. This causes the victim's machine to crash once it receives packets that exploit this vulnerability. In this paper, we are not interested in this type of attack, because, once the vulnerability is detected and patched, the victim is immune to such attacks. Second, there are attacks that exploit a protocol design vulnerability.[1] Such attacks can be fixed by modifying the existing protocols, and by having firewalls check for adherence to protocol specifications; we are not concerned with this type of attack either. The last type of DDoS attacks aim at resource consumption. They exhaust critical resources in the victim's system such as CPU time, memory or network bandwidth, thus causing the disruption of legitimate service. In this paper, we are concerned with a DDoS attack on network bandwidth, also called flooding attack. A flooding attack is very easy to launch as it only requires sending a certain amount of traffic that overwhelms the link connecting the victim to the Internet. [2] An example of flooding attack is shown in Fig. 1. A victim (V) is connected to the Internet through ISP-V, using an access link with bandwidth $C$. The victim is under a DDoS attack from several attack sources hosted by other ISPs, such as ISP-A. The total traffic coming from those sources exceeds the total capacity $C$.

### 2.2 Filtering
Filtering is one of the mechanisms that can help to mitigate DDoS attacks and stop the unwanted traffic from reaching the victim and consuming network bandwidth along the way. For example, in Fig.1, the victim can send a filtering request to its own ISP-V to block all traffic from ISP-A to the victim. ISP-V responds by placing filters at appropriately chosen gateway(s), e.g. GW-V or GW-B. In this paper, we are not concerned with choosing the best gateway within an ISP for placing the filters; instead we look at a single gateway, say GW-V, and how to allocate filters to attackers or attack domains.

---

[1] An example is the well-known TCP three way handshake protocol, which usually allocates resources on the server side in the second step to the handshake. An attacker launches a large number of uncompleted handshakes and overloads the victim's connection tables.

[2] It could also be a problem to ISPs or networks on the way of the attack, but we are not concerned with these cases because they generally are over provisioned

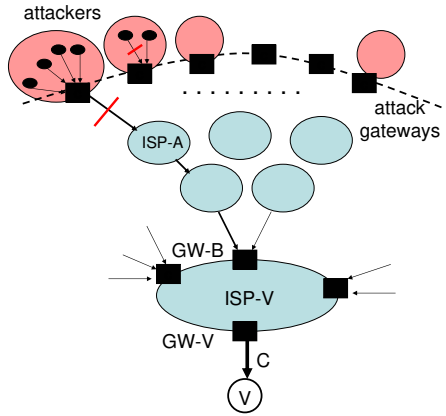

**Figure 1:** The victim (V) is connected to its ISP (ISP-V) through an access router (GW-V) and an access link (with bandwidth $C$). GW-B is a border router of ISP-V. Attackers are located in various ASes behind the attack gateways; the total traffic exceeds the capacity $C$.

By "filters", we refer to access control lists (ACLs), which allow a router to match a packet header against rules. E.g. in the DDoS case described above, the router checks if the packet is going to victim $V$ and coming from attacking host $A$; or the router might check the source IP address and filter out any packet coming from the entire ISP-A. Packet filters in routers are a scarce, expensive resource because they are stored in the expensive TCAM (ternary content addressable memory). A router linecard or supervisor-engine card typically supports a single TCAM chip with tens of thousands of entries. So, depending on how an ISP connects its clients to its network, each client can typically claim from a few hundred to a few thousand filters – not enough to block the attacks observed today and not nearly enough to block the attacks expected in the near future. We formulate two filtering problems: the *single-tier* and the *two-tier* filtering, depending on the granularity of packet filtering (or equivalently, the levels of the attack graph considered). In the single-tier case, we are interested in filtering entire attack gateways, a task for which there are enough filters today; in this context, we seek to filter out traffic so that the total traffic arriving at the victim is below the available bandwidth, while maximizing the preserved legitimate traffic. In the two-tier problem, we are interested in filtering not only attack gateways but also individual attackers, a task for which there are not enough filters in a single router today; the number of filters becomes then an additional constraint.

### 2.3 Related Work
The DDoS problem has been an active research area in recent years and different aspects have been addressed; a taxonomy of attacks and defense mechanisms can be found in [2]. Here we review only aspects related to our work.

Our work relies on existing mechanisms to be able to identify the attack traffic, distinguish it from the legitimate traffic and trace its approximate path back to the attack source [3]. The main difficulty in path identification lies in dealing with source IP address spoofing. [3] Other mechanisms for path identification and traceback include probabilistically sending ICMP messages [5]; mechanisms based on hashing [6] or packet marking [7].

In this work, we focus on filtering at a single router, typically at the victim's gateway. Looking at the bigger picture, several mechanisms have been proposed to enable filter propagation as close to the attack source a possible. For example, Pushback [8] enables routers to propagate filtering upstream hop-by-hop, at the router-level. AITF [9] proposes to communicate filtering information from the victim upstream towards the attack domain, but at the granularity of AS, as opposed to router.

Filtering is not the only mechanism for mitigation of DDoS attacks. Some of the proposed approaches revisit the basic assumption of the Internet architecture, stating that every host can send to any other host, without requiring permission of the receiving host. For example, capabilities which propose that tokens are obtained before establishing a connection with a destination, and that these token are included in each packet [10][11] [12] This proposal requires changing the routers on the Internet and adding new servers and changes the whole Internet architecture. Other proposals use overlay mechanisms to implement a similar concept which is to restrict communication to the victim only through some known well provisioned overlay nodes which can filter and detect attacks [13] [14].

Using filtering could provide a quick solution or first line of defense to DoS attacks, until a permanent one is developed and is already used today in commercially available systems and anti-DoS services [15]. The downside of filtering is that its performance heavily depends on being able to identify attack traffic and distinguish it from legitimate traffic, which is not an easy task. However, it is an available, reactive mechanism that can be used in conjunction with other approaches.

Finally, in this paper, we rely upon data from analysis of worms, to construct realistic attack scenarios. Internet worms are older than DDoS attacks, but are relevant for studying such attacks because they are used as a tool to infect and compromise hosts on the Internet with the attack clients. The Code Red [16] worm is one well-known worm from 2001, which contained code to launch a DoS attack on the website (www1.whitehouse.gov), which did not succeed. Recently, several other worms have been caused huge financial losses, such as Slammer [17], MyDoom, Flash worms [18] and others and have attracted a lot of researcher's attention. Prolexic [15] is also regularly publishing a very informative "Zombie Report", on the most infected hosts per country, network service-provider and other meaningful groupings [19]; these data are collected from actually launched DDoS attacks. A detailed characterization of DDoS attacks is much needed as input for testing and evaluating the effectiveness of proposed solutions.

## 3. FORMULATION OF OPTIMAL ALLOCATION OF FILTERS

### 3.1 General Discussion

In general, one might consider allocating filters at any level of the attack graph, see Fig.1. There is clearly a trade off between filtering granularity (to maximize goodput) and the number of filters. If there were no constraints on the number of filters, the maximum throughput of good traffic (goodput) would be achieved by allocating filters as close to individual attackers as possible. The gateway in question (GW-V) faces the following tradeoff. Ideally, GW-V would like to filter out all attackers and allow all good traffic to reach the victim. Unfortunately, in a typical DDoS attack, there are not enough filters to individually filter all IP addresses of attack hosts. A solution is to aggregate attack sources into a single filter; in practice, there are enough filters available to filter at that granularity. E.g. GW-V could summarize several attack sources coming from the same domain, e.g. behind GW-1, into a single rule and filter out the entire domains, as shown in Fig. 2. However, there is also legitimate traffic coming from each domain. Therefore, filtering at the granularity of attack gateway-tier causes "collateral" damage to legitimate traffic that falls into the range of the IP addresses described by the filter. This problem, referred to as the "single-tier filtering", is studied in section 3.2 so as to preserve the maximum amount of legitimate traffic while meeting the capacity constraint. This turns out to be a knapsack problem that can be solved by a greedy algorithm (shown in Algorithm 1).

In practice, there are more filters ($F$) than attack gateways ($N < F$), but less filters than individual attackers ($F < \sum_{i=1}^{N} M_i$) (see Fig. 3). Filtering at the gateway level is feasible but causes the collateral damage discussed above, due to its coarse granularity. Filtering at the attacker's level would preserve the maximum possible throughput but it is not realistic (due to the high number of attackers as well as due to spoofing); we still consider it as an upper bound for performance. A practical and effective compromise between the two extremes can be the two-tier filtering, shown in Fig. 3. In the two-tier filtering, we can choose to filter either at gateways' granularity (e.g. filter 1 in Fig. 3) or at attackers' granularity (e.g. filter 2 in Fig. 3). The optimal allocation of filters to individual attack sources, or to entire attack gateways, depends on the characteristics of the attack (distribution and intensity) as well as on the number of available filters. Furthermore, the successful containment of the DDoS attack crucially depends on the optimization of the filter allocation.

### 3.2 Single-Tier Allocation

The single-tier scenario is shown in Fig.2. There are $N$ attacking gateways, each generating both good ($G_i$) and bad ($B_i$) traffic toward the victim; the total traffic toward the victim exceeds its capacity $C$. Gateway GW-V allocates filters to block the attack traffic towards V. There are enough filters to allocate to the $N$ gateways. The objective is to allocate filters to limit the total traffic below the available capacity, so as to maximize the amount of legitimate traffic that is getting through to the victim (because this is what the victim cares about, e.g. revenue for a web server).

---
[3]Because ingress filtering [4] is not widely adopted, it is easy to spoof the source IP address of any packet. However, most attack tools do not need to spoof, because of the large number of attackers, it is useless to spoof.

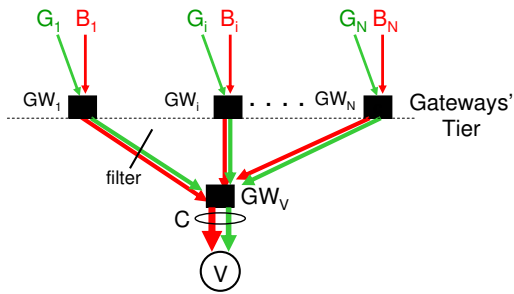

Figure 2: Single-Tier Filtering Problem

Let us use $x_i = 1$ and $x_i = 0$ to indicate whether we allow or block all traffic from gateway $i$. The problem of optimal allocation of filters is to choose $\{x_i\}_i^N$:

$$max \ \frac{\sum G_i \cdot x_i}{\sum (G_i + B_i) \cdot x_i}$$
$$s.t. \sum_{i=1}^{i=N} (G_i + B_i) \cdot x_i \leq C \quad (1)$$
$$and \ x_i \in \{0, 1\}, i = 1, 2, ..N$$

We noticed that the filter allocation problem is essentially a 0-1 knapsack problem [21]. Recall that in the knapsack problem, we choose some among $N$ objects, each with profit $v_i$ and a weight $w_i$, so as to maximize the total profit, subject to a total weight constraint. In our case, the objects are the attacking nodes with profits and weights $G_i$ and $G_i + B_i$ respectively; and there is a constraint $C$ on the victim's bandwidth. This is well-known to be a computationally hard problem. However, we need computationally efficient solutions, because the filter allocation should be decided in real-time during the attack.

The continuous relaxation of the problem (where $x$ is no longer binary, but instead $0 \leq x_i \leq 1$) can be interpreted as placing rate-limiters: we allow ratio $x_i$ of the traffic coming from node $i$ to get to the victim. This corresponds to the fractional knapsack problem, which can be solved optimally using a greedy algorithm [21]. The algorithm in Algorithm 1, shown below, sorts nodes in a decreasing order of efficiency $\frac{G_j}{G_j+B_j}$,[4] and greedily accepts ($x_i = 1$) nodes with the maximum efficiency, until a critical node $c$, which if allowed will exceed the capacity. Traffic from all remaining nodes is filtered out ($x_i = 0$) and installs a rate-limiter to the critical element ($x_c = \frac{C - \sum_{j=1}^{j=c-1} G_j + B_j}{G_c + B_c}$) to use the remaining capacity. This requires only $O(nlogn)$ steps for sorting and $O(n)$ for filer/rate-limiters allocation.

---

**Algorithm 1** Greedy Algorithm for the Single-Tier.

- Order nodes in decreasing order $\frac{G_j}{G_j+B_j}$. W.l.o.g. $j = 1, 2, ..N$ from largest to smallest efficiency.
- Find the critical node $c$ s.t.: $\sum_{j=1}^{j=c-1} G_j + B_j < C$ and $\sum_{j=1}^{j=c} G_j + B_j > C$
- Allocate filters to nodes $i = 1, 2, ...N$ as follows:
  - $x_j = 1$ for $j = 1, 2, ..c - 1$ (allow to pass)
  - $x_c = \frac{C - \sum_{j=1}^{j=c-1} G_j + B_j}{G_c + B_c}$ (rate limiter)
  - $x_j = 0$ for $j = c + 1, ..n$ (filters)

---

Notice, that it is impractical to allocate rate-limiters to all attacking nodes, because rate-limiters are expensive resources and require keeping state. Fortunately, the optimal solution of the fractional problem turned out to be $(x_1, ...x_{c-1}, x_c, x_{c+1}, ..., x_N) = (1, ...1, x_{c-1}, 0, ...0)$, thus using $C - 1$ filters and exactly one rate-limiter, which matches well current router resources. [5]

### 3.3 Two-Tier Allocation

The two-tier problem is the following. Consider $N$ attack gateways and $M_i$ attack hosts behind attack gateway $i$, i.e. the last two tiers in Fig.1. Each attacker contributes both good ($G_{ij}$) and bad traffic ($B_{ij}$), $i = 1, 2..N, j = 1, 2...M_j$. $x_{ij} \in 0, 1$ depending on whether we allocate a filter to attack-host $j$ behind gateway $i$. $x_i \in 0, 1$ depending on whether we allocate a filter to attack-gateway $i$; if $x_i = 0$, then all traffic originating behind GW-i is blocked, and there is no need to allocate additional filters to attackers $(i, j), j = 1, 2, ...M_i$.

The problem is how to choose $\{x_i\}$'s, $\{x_{ij}\}$'s, given the constraints $C$ on the victim's capacity and on the available number of filters $F$ at the gateway:

$$max \ \sum G_{ij} \cdot x_i \cdot x_{ij}$$
$$s.t. \sum_{i=1}^{i=N} \sum_{j=1}^{j=M_i} (G_{ij} + B_{ij}) \cdot x_i \cdot x_{ij} \leq C \quad (2)$$
$$\sum (1 - x_i) + \sum (1 - x_{ij}) \leq F$$
$$where \ x_i \in \{0, 1\}, \ i = 1, 2, ..N$$
$$and \ x_{ij} \in \{0, 1\}, \ j = 1, 2..., M_j$$

The two-tier problem is harder than the single-tier one: it is a variation of the cardinality-constrained knapsack [21], and the optimal solution (in $O(NMF)$) cannot be found efficiently. We formulate the problem using dynamic programming and obtain its optimum solution as a base line for comparison, but we point out that the dynamic programming algorithm is computationally very expensive and can not be used in real time.

---

[4] Technically, maximizing $\sum \frac{G_i}{G_i+B_i}$ is not the same as maximizing $\sum G_i$. However because the optimal solution has $\sum G_i + B_i \simeq C$, it is the same in practice.

[5] If only filters are allowed, we can choose $x_c = 0$ so that the total traffic does not violate the capacity constraint of the congested link.

*Definitions.* Consider the two-tiers configurations, shown in Fig. 3. There are $N$ gateways. A gateway $n$ generates legitimate traffic $G_n$ and also attack traffic from $M_n$ attack sources; w.l.o.g. consider that the attacker sources are ordered from worst to best: $b(n,1) > ... > b(n, M_n)$. Therefore, each gateway generates total traffic $C_n = G_n + \sum_{i=1}^{M_n} b(n,i)$. Before filtering, the total traffic exceeds the victim's access bandwidth (capacity) $C$: $\sum_{i=1}^{N} C_n > C$. We are interested in placing $F$ filters across the $N$ gateways, so as to bring the total traffic below $C$, while maximizing the total goodput after filtering $T_N(C, F)$. $T_N^*(C, F)$, can be computed recursively as summarized in Algorithm 2. The recursion proceeds considering one more gateway at a time; the order in which gateways are considered is not important. Let $T_i^*(c, f)$, for $i \leq N$, be the maximum goodput of the smaller problem, i.e. with optimal placement of $f \leq F$ filters considering only gateways $\{1, 2, ..i\}$ and capacity up to $c \leq C$. Assume that, in previous steps, we have already obtained and stored the optimal solutions $T_i(c, f)$ considering only gateways $1, 2, ...n-1$, for all values of $c = 0, 1, ..C$ and $f = 0, 1, ...F$. Then $T_n^*(c, f)$ can be computed from the Bellman recursive equation:

$$T_n^*(c,f) = \max_{x=0,1,..f} T_{n-1}^*(c - (C_n - \sum_{j=0}^{j=x} b(n,j)), f-x) + G_n \quad (3)$$

This recursive equation is at the heart of Alg.2 (see line 23).

*Intuition.* In step $n$, we consider gateway $n$ together with the previous gateways $1, 2, ...n-1$. The $f$ available filters can be split among two groups of gateways: $\{1, 2, ..n-1\}$

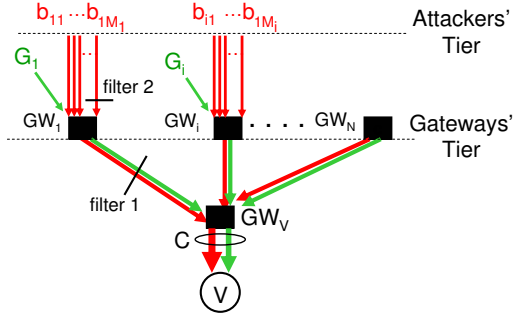

Figure 3: Two-Tiers Filtering Problem

and $\{n\}$; $x \leq f$ filters are assigned to the new gateway $n$ and the remaining $f - x$ filters are assigned to the previous gateways $\{1, 2, ..n - 1\}$. The $x$ filters assigned to $GW_n$ are used to block the $x$ worst attackers. Therefore, $\sum_{j=0}^{j=x} b(n,x)$ bad traffic is blocked and the remaining $C(n) - \sum_{j=0}^{j=x} b(n,x) = G_n + \sum_{j=x+1}^{j=M_n} b(n,x)$ traffic goes through ($gwn_{unfiltered}$ in line 24), consuming part of the total capacity $c$. The remaining $f - x$ filters are optimally assigned to gateways $1, 2, ...n - 1$. Recall that we have previously obtained and stored the optimal solutions $T_{n-1}^*(c, f)$ considering only gateways $\{1, 2, ...n - 1\}$, for all $c$ and $f$; therefore, we already know the best allocation of $f - x$ filters across gateways $\{1, 2, ...n - 1\}$ so as to get the maximum goodput $T_{n-1}^*(c - (C(n) - \sum_{j=0}^{j=x} b(n,x)), f - x)$.

We consider all possible values of $x$ and choose the value among $0 \leq x \leq f$ that gives the maximum goodput (line 33 in Alg.2). There are some values of $x$ that deserve special attention:

- $x = 0$ means that we assign no filters to gateway $n$, in which case our best goodput is the same as before, enhanced by the goodput of the current gateway: $T_{n-1}^*(c - C_n, f) + G_n$ ($max0$ in line 12 of Alg. 2).

- $x = 1$ means that we assign exactly one filter to gateway $n$, either at attacker or at gateway level. If we assign this one filter to an attacker, it should be the worst attacker $b(n, 1)$ (line 16 in Alg.2). If this one filter is assigned to the entire gateway, then the entire traffic $C_n$ from gateway $n$ is filtered out and all goodput comes from the previous gateways $T_{n-1}^*(c, f - 1)$ (see line 18 of Alg.2). We need to compare the two possibilities and choose the one that maximizes the overall goodput ($max1$ in line 19 of Alg.2).

- We consider increasing values of $x$ until we either run out of filters ($x = f$) or we filter out all attackers in this gateway ($x = M_n$). Therefore, $x$ can increase up to $min\{f, M_n\}$ (line 23 in Alg. 2).

Other technicalities in Algorithm 2 include the obvious initializations (lines 1-3) and assigning $T^* = 0$ to infeasible problems (line 3-$2^{nd}$ case and line 28).

*Optimal Substructure.* The reason that we are able to compute the optimal solution using dynamic programming (DP) is because the problem exhibits the optimal substructure property.

**Proposition.** If $a^*$ is the optimal solution for problem $(n, c, f)$, then it contains a part $a_{\{1,...n-1\}}^* \subset a^*$ (corresponding to the filters assigned to the first n-1 gateways) which must also be the optimal solution for the smaller problem $(n - 1, C - (C_n - \sum_{j=0}^{j=x} b(n, x)), f - x)$.

PROOF. $a^*$ is the optimal solution for problem $(n, c, f)$, achieving maximum goodput $T_n^*(c, f)$.[6] This solution (filter

---

[6] $a^*$ will have the form of a vector $(1, 0, 0, ..., 0, 1)$; 0/1 describes whether an attacker or gateway has been filtered out or not; the attackers and gateways should be listed in the same order they are considered in the DP.

**Algorithm 2** Dynamic Programming (DP) Formulation for the Two-Tiers Filtering Problem

1: **for** $n = 1, 2, ..N$ **do**
2:    $T_n^*(c=0,:) = 0$
3:    $T_n^*(:,f=0) = \begin{cases} \sum_{n=1}^{N} G_n & \text{if } \sum_{n=1}^{N} G_n < C \\ 0 & \text{otherwise} \end{cases}$
4: **end for**
5:
6: **for** $n \in [1, N]$ **do**
7:   **for** $c \in [1, C]$ **do**
8:     **for** $f \in [1, F]$ **do**
9:       /* $x$ out of $f$ filters are assigned to $GW_n$ */
10:
11:       /* assign $x = 0$ filters to $GW_n$*/
12:       $max0 = T_{n-1}^*(c - C_n, f) + G_n$
13:
14:       /*assign $x = 1$ filter to $GW_n$*/
15:       /* ...either at gateway level*/
16:       $max1_{gw} = T_{n-1}^*(c, f-1)$
17:       /* ...or at attacker level*/
18:       $max1_{att} = T_{n-1}^*(c - (C_n - b(n,1)), f-1) + G_n$
19:       $max1 = max\{max1_{gw}, max1_{att}\}$
20:       $max = max\{max0, max1\}$
21:
22:       /* assign $x \geq 2$ filters at attack level. */
23:       **for** $x \in [2, min(f, M_n)]$ **do**
24:         $gwn_{unfiltered} := C_n - \sum_{j=1}^{x} b(n,j)$
25:         **if** $c > gwn_{unfiltered}$ **then**
26:           $temp := T_{n-1}*(c - gwn_{unfiltered}, f - x) + G_n$
27:         **else**
28:           $temp := 0$
29:         **end if**
30:         **if** $temp > max$ **then**
31:           $max := temp$
32:         **end if**
33:         $T_n^*(c, f) := max$
34:       **end for**
35:     **end for**
36:   **end for**
37: **end for**

assignment) must have two parts $a^* = (a_{\{1,2..n-1\}}^*, a_{\{n\}}^*)$. The first part, $a_{\{1,2..n-1\}}^*$ [7], describes how filters are placed across gateways $\{1, 2, ..n - 1\}$. The second part, $a_{\{n\}}$ [8] describes how filters are assigned to gateway $\{n\}$ only. Let's look carefully at the optimal solution $a^*$: it assigns some number of filters ($x$) to gateway $n$ and the remaining ones ($f-x$) to gateways $\{0, 1, ..n-1\}$. This means that $\sum_{j=0}^{j=x} b(n, x)$ out of $C_n$ traffic is filtered out at gateway $n$ and the remaining $C_n - \sum_{j=0}^{j=x} b(n, x)$ is left unfiltered. In summary, the optimal solution can be partitioned in two parts $a^* = (a_{\{1,2..n-1\}}^*, a_{\{n\}}^*)$ that contribute to the maximum throughput:

$$T_n^*(c, f) := T\big|_{a^*} = T\big|_{a_{\{1,2..n-1\}}^*} + T\big|_{a_{\{n\}}^*}$$

Assume that $b$, and not $a_{\{1,2..n-1\}}^*$, is the optimal filter as-signment for the smaller problem $(n-1, C-(C_n-\sum_{j=0}^{j=x} b(n,x)), f-x)$. Then, by definition, of the optimal filtering, it achieves larger goodput than the substructure $a_{\{1,2..n-1\}}^* \subset a^*$: $T_{n-1}^* := T\big|_b > T\big|_{a_{\{1,2..n-1\}}^*}$.

---
[7] a vector $(1, 0, 0, ...)$ starting from the left of vector $a^*$
[8] the remaining part at the right side of vector $a^*$

Now, we can construct another solution $d$ for the larger problem $(n, c, f)$ as follows. Replace the first part $a_{\{1,2..n-1\}}$ of $a^*$ with $b$, for assigning $f - x$ filters up to gateway $n - 1$, which would fit within capacity $C - (C_n - \sum_{j=0}^{j=x} b(n,x))$. Then, do exactly the same assignment as the DP would do, in Eq. 3, for assigning the $x$ remaining filters to gateway $n$. This newly constructed filter assignment $d$ has two parts $d = (b, d_{\{n\}})$ that contribute to the total goodput.

The first part $b$ is over gateways $\{1, 2, ..n - 1\}$. We constructed this part to be the same as the optimal assignment of $f - x$ filters over gateways $\{1, 2, ..n - 1\}$, with available capacity $C - (C_n - \sum_{j=0}^{j=x} b(n,x))$. Therefore it achieves optimal goodput $T\big|_b := T_{n-1}^* \geq T\big|_{a_{\{1,2..n-1\}}^*}$.

The second part $d_2$ is an assignment over only gateway $\{n\}$. We constructed it to do exactly what the DP would do at step $n$ with $x$ available filters: either filter out the worst $x$ attackers of gateway $n$ (i.e. attackers $b(n, 1)...b(1, x)$) or filter out the entire gateway (if $x = 1$ is assigned at gateway level). Therefore, $d_2$ is by construction is the exact same assignment as the DP's: $d_2 = a_{\{n\}*}$ and results in the same goodput: $T\big|_{d_2} = T\big|_{a_{\{n\}}^*}$.

Therefore, we constructed a solution $d = (b, d_2)$ which performs better than the DP solution $a*$:

$$T\big|_d = T\big|_b + T\big|_{d_{\{n\}}} > T\big|_{a_{\{1,2..n-1\}}^*} + T\big|_{a_{\{n\}}^*} = T\big|_{a^*} \quad (4)$$

This is a contradiction because we assumed that $a*$ was an optimal solution for the bigger problem $(n, c, f)$. Therefore the substructure $a_{\{1,...n-1\}}^*$ of the optimal solution $a^*$ has to be the optimal solution for the smaller problem. $\square$

*Cost of the Dynamic Programming.* Computing the optimal solution value $T_n^*(C, F)$ using dynamic programming (DP) can be seen as filling up a table of size $N \cdot C \cdot F$. For realistic large values of $N$, $C$, $F$ this can be prohibitively large, both from a run-time and from a memory point of view. While the number of gateways $N$ can be moderate, the capacity $C$ (normalized in units of the smallest attack rate) and $F$ (in the order of thousands or tens of thousands) can be quite large in practice. An idea might be to work with coarser increments of $C$ and $F$ - which brings us already in the realm of heuristics for the DP, which we need to explore anyway. Nevertheless, computing the optimal solution is still important as a benchmark for evaluation of any proposed heuristic.

### 3.3.1 Properties of the Optimal Solution
From the simulations of the dynamic programming, in section 4.4, we made the following preliminary observations on the behavior of the optimal solution.

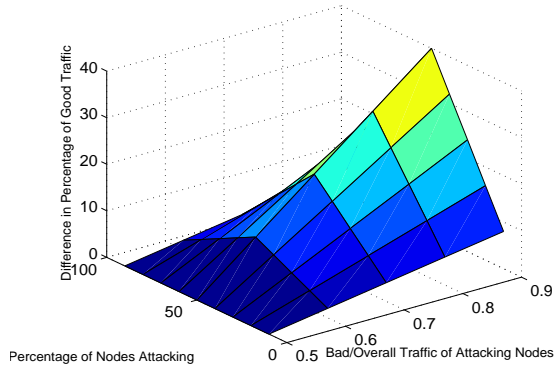

**Figure 4:** Improvement from using optimal filtering for various attack intensities (% of attacking nodes, $B/(G+B)$). We consider $n = 1000$ attacking nodes, all sending at the same rate (10Mbps).

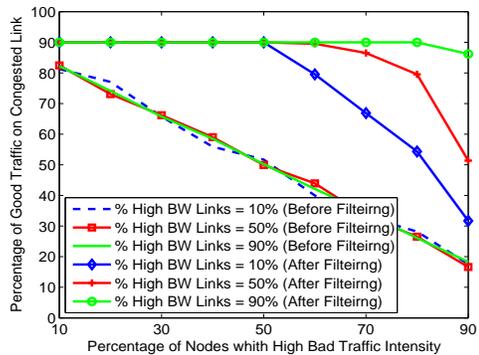

**Figure 5:** Improvement from using optimal filtering for various % of nodes with high intensity, and various % of nodes sending at higher rate (100Mbps). $H = 0.9$ is fixed for all attacking nodes.

*Comparison to single-tier filtering.* Let $T_N(C, f)$, $G_N(C, f)$, $A_N(C, f)$ be the maximum goodput achieved by the optimal placement of $f$ filters across $N$ gateways, considering multi-tier placement, single-tier placement at gateways and single attackers respectively. Given the same attack scenario:

- $T_N(C, f) \geq A_N(C, f)$ and $T_N(C, f) \geq G_N(C, f)$. This is expected, because by definition, the optimal solution of the multi-tier problem considers placing all filters at gateway and attack level, as special cases.

- $G^* < T < A^*$ where: $G^*$ is the optimal filtering for single gateway tier and $A^*$ is the optimal filtering for the single attack level filtering. Single tier filtering does not have a constraint on the number of filters and is only constrained by the "collateral damage" on legitimate traffic.

- As $f \uparrow$, $T$ converges to $A^*$, the optimal solution for attacker's single tier, without a constraint on the number of filters.

*A preliminary note.* The core tradeoff when we consider filtering a single gateway is whether we should filter it out entirely (thus filtering out both $G_i$ and $B_i$) or we should use a certain number of filters $f$ at attack-tier. How many filters, out of our total budget, does it worth spending on a gateway with traffic $(G_i, b(i, 1), ...(b, n))$? Looking at the structure of the optimal solution, it seems to follow a threshold rule for deciding whether to filter out an entire gateway or not. This threshold depends on the attack distribution and on the number of available filters. We are currently working on formalizing this informal, but intuitive, observation.

## 4. SIMULATIONS
## 4.1 Single-Tier Artificially Generated Scenarios

We systematically considered a wide range of scenarios and here we showcase some representative results. Let us fix the number $N$ of attack nodes; we considered $N = 10, 100, 1000$.

Let us control the intensity of the attack through a simple model with three parameters. (i) the bandwidth at which each node sends is a configurable parameter. (ii) x% of the nodes that are attacking and the remaining (100-x)% send legitimate traffic (iii) attacking nodes have all the same bad-to-overall traffic ratio $H = \frac{B}{B+G}$; the legitimate nodes have ratio $1 - H$ of bad to overall.

Fig.4 shows the results for $N = 1000$ nodes, which all send at the same rate (10Mbps). We consider all combinations of $x \in \{0, 100\}\%$ and $H \in (0.5, 0.9)$ and we look at the difference in the % of good traffic on the congested link, before and after optimal filtering. The figure shows that there is always improvement, with the best improvement (40%) achieved when 50% of all nodes are attackers, sending at $H = \frac{B}{B+G} = 0.9$.

Then, we also vary the sending rate of each node. We randomly pick 10%, 50% or 90% of the nodes to have 10 times more bandwidth than the rest (i.e. 100Mbps). The reason we look at heterogeneous bandwidths is that a node should be filtered based not only on the ratio $\frac{B}{B+G}$, but also on its total contribution $B + G$ to the capacity of the congested link. Fig.5, shows that optimal filtering significantly helps in this case.

### 4.1.1 Varying the number of attacking nodes
In this section, we increase the number of nodes and we are interested not only in the % of good traffic preserved, but also in the number of filters required. We compare *optimal filtering* to 3 benchmark policies:

- *Uniform rate limiting*: rate-limit all nodes by $\frac{C}{total\ traffic}$, to make sure the total traffic does not exceed the capacity. Notice, that this policy is equivalent to *no filtering* in terms of percentage of good to overall traffic on the congested link.

- *Random filtering*: randomly place the same number of filters as the optimal policy.

- *Max-min rate limiting*: admit the low-rate nodes first while allocating the same bandwidth to the high rate

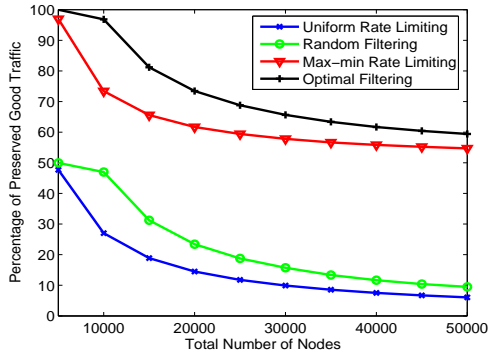

(a) % Good Traffic Preserved after Filtering

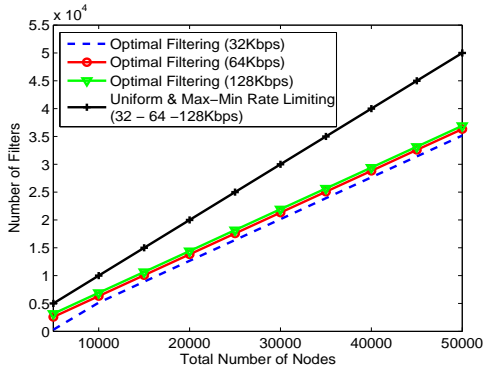

(b) Number of filters used.

**Figure 6: Performance of Optimal Filtering for the attackers' one tier.**

ones; then distribute the excess capacity fairly among the unsatisfied remaining nodes.[9]

We vary the number of attackers (from 1000 to 50000) and we allocate filters to individual attackers (attackers' one-tier problem )[10]. In Fig6, optimal filtering clearly outperforms the other policies: it preserves more good traffic using less filters. However, the number of filters increases linearly with the number of attackers, which clearly does not scale for a large number of attackers.

To deal with this scalability issue, we solve the one-tier problem at the gateway level. We consider again an increasing number of attackers (from 1000 to 50000), but this time attackers are evenly spread behind $n = 1000$ gateways (as in Fig. 1) and we allocate filters to gateways, not to individual attackers. The results are shown in Fig. 7. The optimal policy again outperforms the others: it preserves significantly more good traffic while using much less filters. However, there are several differences from filtering at the attackers' tier, all due to the coarser filtering granularity. First, we need less filters, but the % of preserved traffic drops below 50% in the case of larger number of attackers. Second, the number of filters used by the optimal policy increases fast up to around 90% and then saturates, because otherwise all traffic would be blocked. Third, the max-min policy performs much worse now; also from a practical point of view, the uniform and max-min policies are less attractive, because they use rate-limiters on all nodes, which is unrealistic.

---
[9]The motivation is to do a more informed rate limiting than the uniform rate-limiting, and also to be friendly to the nodes that emit at lower rates.

[10]In this simulation scenario, we vary $N$, but we make sure that the total good traffic is below the capacity (in particular $\sum_{1}^{N} G_i = \frac{C}{2}$), because this is the practical case of adequate provisioning. To construct such an assignment, we assign $\frac{C}{2}$ over half of the nodes assigning $\frac{C}{N}$ to every other node and we randomly pick $N/2$ nodes and assign them bad traffic. We make sure that the total traffic emitted by each node is no more than its maximum rate (32, 64, or 128 kbps)

**Table 1: Code-Red Scenarios**

| Country | GW | Code Red I | | Code Red II | |
|---|---|---|---|---|---|
| | | % of Good Traffic from [20] | % of Bad Traffic from [16] | % of Good Traffic | % of Bad Traffic |
| USA | 1 | 36.27 | 43.9 | 36.2 | 45.9 |
| Korea | 2 | 5.8 | 11.5 | 0 | 12 |
| China | 3 | 18.35 | 10.3 | 24.1 | 0 |
| Taiwan | 4 | 2.46 | 6.1 | 2.4 | 16.7 |
| Canada | 5 | 3.64 | 5.4 | 3.6 | 5.4 |
| UK | 6 | 6.74 | 5.2 | 6.7 | 5.3 |
| Germany | 7 | 8.4 | 5.1 | 8.4 | 5.2 |
| Australia | 8 | 2.5 | 4.3 | 2.5 | 1.1 |
| Japan | 9 | 13.91 | 4.2 | 14.2 | 0 |
| Netherlands | 10 | 1.93 | 4.1 | 1.9 | 8.4 |

### 4.2 Realistic Attack Scenarios

First, we used data from the analysis of two recent worms, Code-Red [16] and Slammer [17] to construct realistic attack distributions as in the single-tier section. Another source of data we used for the attack traffic distribution is Zombie Report [19] published by Prolexic [15]. This report contains the percentage of bots, grouped per country, network, ISP and other meaningful groupings; we use the data referring to the number of infected hosts per country. We assume that if a victim is under attack that traffic would come from ten countries. We consider the ten first countries and assume that they are behind ten different gateways[11]. The distribution of attack traffic for the Code-Red, Slammer Zombie scenario is summarized in the last column of Tables 1, 2 and 3.

We consider a typical victim – a web-server with 100Mbps access link. We also consider that each country is in a different AS, thus is behind a different gateway; we then use the number of attack sources per gateway, as reported in [16, 17, 19] [12] and shown in the $4^{th}$ column of Table 1. For the legitimate traffic, we use the breakdown of Internet users per country reported in [20] and shown in the $3^{rd}$ column of Table 1, 2 and ??. We consider that both attackers and legitimate users send at the same rate (32kbps, 64kbps or 128kbps), corresponding to upstream dialup/dsl. Therefore, if the total number of legitimate users is $N$ and that of attackers is $M$, then the amount of good and bad traffic coming from gateway $i$ is $G_i = N \cdot$ (% users behind gateway $i$) $\cdot$ (rate) and $B_i = M \cdot$ (% users behind gateway $i$) $\cdot$ (rate). Our rationale is that the number of legitimate users is representative of the legitimate traffic coming from each country. We use the number/percentage of legitimate users, to compute the % of total good traffic generated by each gateway. The result is summarized in the third column in each attack scenario.

---
[11]note that the first ten countries sum up to be 65% so we distribute the rest of the 35% of the attack traffic uniformly on these countries

[12]In [16] 80% of the total attack comes from 10 countries; we distributed the rest 20% of the attack uniformly across the lower 8 countries.

**Table 2: Slammer Scenario: attack launched by a population of hosts infected by a worm similar to Slammer.**

| Country | GW | % Good Traffic | % Bad Traffic |
|---|---|---|---|
| USA | 1 | 36.3% | 44.6% |
| South Korea | 2 | 5.8% | 13.6% |
| China | 3 | 18.5% | 8% |
| Taiwan | 4 | 2.4% | 5.7% |
| Canada | 5 | 3.6% | 4.6% |
| Australia | 6 | 2.5% | 4.2% |
| UK | 7 | 6.7% | 3.8% |
| Japan | 8 | 13.9% | 3.5% |
| Netherlands | 9 | 1.9 % | 3.3% |
| Unknown | 10 | 8.4% | 8.7% |
| Total | | 100% | 100% |

**Table 3: Prolexic Scenario: attack launched by a bots population, similar to the one in the Prolexic Zombie Report.**

| Country | GW | % Good Traffic | % Bad Traffic |
|---|---|---|---|
| US | 1 | 36.5% | 21.5% |
| China | 2 | 18.5% | 14.5% |
| Germany | 3 | 8.5 % | 13.5% |
| UK | 4 | 6.78% | 8.5% |
| France | 5 | 4.59% | 8.5% |
| Brazil | 6 | 4% | 7.5% |
| Japan | 7 | 13.99% | 7.5% |
| Phillippines | 8 | 1.4% | 6.5% |
| Russia | 9 | 13.94% | 6.5% |
| Malaysia | 10 | 1.8% | 5.5% |
| Total | | 100% | 100% |

In summary, we construct 3 realistic attack scenarios using data from [20] for the good traffic and from the analysis of code-red worm, slammer worm and in the Prolexic report for the bad traffic. The attack scenarios for the CodeRed, Slammer and Prolexic are summarized in Tables 1, 2 and 3 respectively.

### 4.3 Results For Single-Tier

We simulated the Code-Red scenario, (*CodeRed I* – columns 3 and 4 of Table 1) for a number of legitimate users $N$ from 1000 to 10000 and attackers $M$ from 1000 to 10000, and we compared the amount of good traffic preserved without any filtering and with optimal filtering. The results are shown in Fig. 8 for 32Kbps sending rate. (The results for 64kbps and 128kbps show similar trends and are omitted here).

Fig. 8(a) shows the % of good traffic preserved. When the

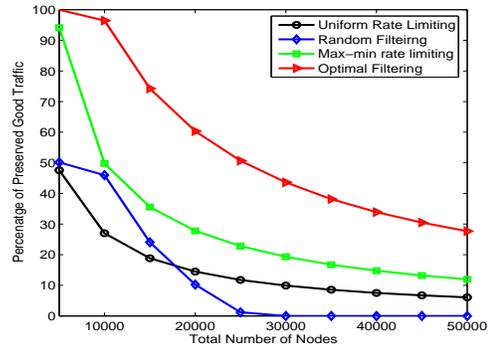

(a) % Good Traffic Preserved after Filtering

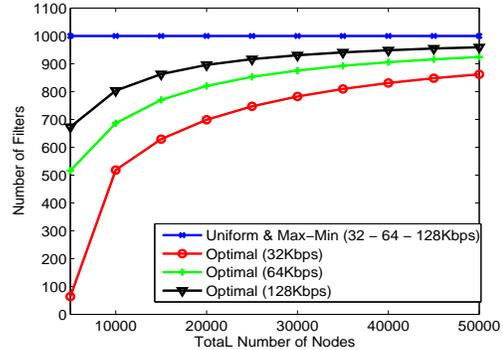

(b) Number of filters used.

**Figure 7: Performance of Optimal Filtering for the gateways' one tier (1000 gateways, same number of attackers behind each gateway).**

total good traffic is less than the capacity of the congested link,[13] and the number of attackers was between 1000 and 2000, optimal filtering preserves 100% of the good traffic. As the number of attackers increases, the % of good traffic preserved drops; e.g. for 1000 users and 10000 attackers, optimal filtering preserves 55% of the good traffic. This is because filtering at the gateway level is as if it is based on destination address or source address with granularity of domains; better results could be achieved if a finer granularity of filtering could be applied (i.e. source address of individual attackers), as in the multi-tier case later.

Fig. 8(b) shows the % of the capacity of the congested link that consists of good traffic. When the number of attackers is comparable to the number of legitimate users (e.g. 10000 attackers and 10000 users), we observe that the optimal filtering preserves 25% more good traffic, which increases the percentage of good traffic on the congested link from 50% to 75%. However, at the extremes where the number of legitimate users is much smaller (e.g. 1000 users and 10000 attackers) or much larger (e.g. 10000 users and 1000 attack-

---

[13]When the good traffic exceeds the capacity, we cannot preserve 100% of good traffic any way. This can be seen as a combination of a flash-crowd and DoS attack. In the rest of the paper, we focus on cases where the good traffic does not exceed the capacity of the link (which is the case with normal operation and good provisioning).

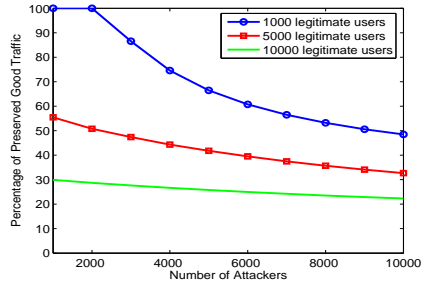

(a) % Good Traffic Preserved after Optimal Filtering

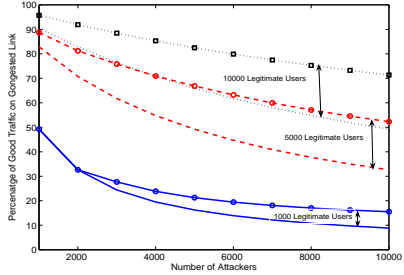

(b) Good % Link BW before/after Optimal Filtering.

**Figure 8: Performance of Optimal Filtering for scenario Code Red I.**

ers) than the number of attackers then the optimal filtering increases the preserved good traffic only marginally ( 10%).

The improvement achieved above was moderate, because all gateways had both good and bad traffic. If there are gateways that carry only good or bad traffic, then filtering would be able to better separate good from bad traffic and further improve performance. To demonstrate this, we modify the distributions of good and bad traffic per gateway, as shown in scenario *Code Red II* – $5^{th}$ and $6^{th}$ column of Table 1. This modification maps to real life scenarios in which a certain website has only customers in some, but not all, countries (ASs). Also it is reasonable to assume that the attacker will not be able to compromise hosts in ASes that span all countries, thus there are some gateways with only bad or good traffic.

We used the same simulation setup as for the *Code-Red I* scenario and the results are shown in Fig. 9. The trends are similar but the improvement is more substantial: the % of the capacity of the congested link used by good traffic improves up to 50% with optimal filtering. In the case of 10000 legitimate users, optimal filtering allows only good traffic through the congested link until the capacity is used. The same behavior can be observed for 5000 legitimate users with 64Kbps and 128Kbps rates.

## 4.4 Results For Two-Tier

Figures 10, 11, and 12 show the performance of the optimal two-tier filtering for the Code-Red scenario, the Slammer scenario and the Zombie scenario respectively. In all three

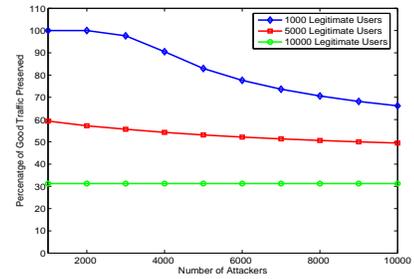

(a) % Good Traffic Preserved after Optimal Filtering

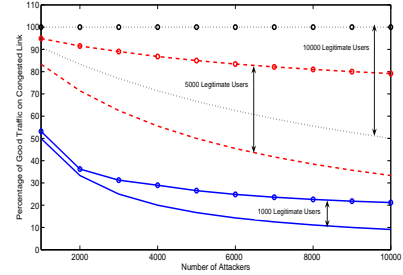

(b) Good % Link BW before/after Optimal Filtering.

**Figure 9: Performance of Optimal Filtering for scenario Code Red II.**

cases, we increase the number of attackers[14] and we look at how well filtering can handle the increasing attack traffic. The performance metrics of interest are (a) the % goodput preserved after filtering and (b) the number of filters used in the process. As a baseline for comparison, we also show the performance of the optimal single-tier filtering at gateway and attack level.

In all three figures, the optimal solution performs very similarly; although they are based on different distributions of good and bad traffic. As expected, filtering at attackers' level (plain red line) gives the upper bound for the preserved goodput. Indeed, one can preserved 100 % of the good traffic by filtering out each individual attacker[15] but requires as many filters as the number of attackers, which is not feasible in practice. Filtering at the gateway level (shown in dashed green line) provides a lower bound to the preserved goodput (because it filters out together both good and bad traffic behind the same gateway) but uses a small number of filters. Multi-tier filtering lies in the middle (blue curves in the middle): it provides a graceful degradation of preserved goodput, using only a small number of filters. The larger the number of filters available for multi-tier filtering, the closer the preserved goodput will be to the upper bound.

In the previous simulations, we increased the number of at-

---

[14] Because the computation of the optimal solution through DP was computationally intense, we consider a relatively small number of k and n in order for the simulations to complete.

[15] assuming there are no hosts that produce both good and bad traffic

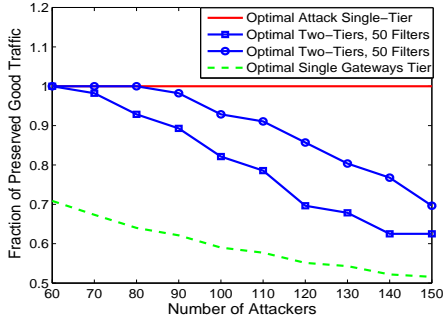

(a) % Good Traffic Preserved

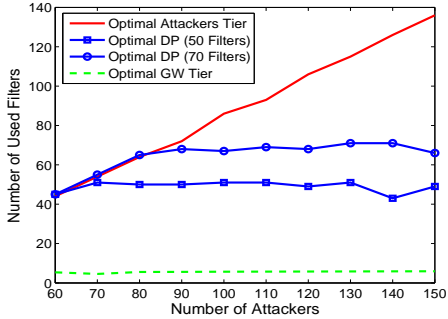

(b) Number of filters used.

**Figure 10: Performance of the Optimal Multi-tier Filtering for the CodeRed scenario.**

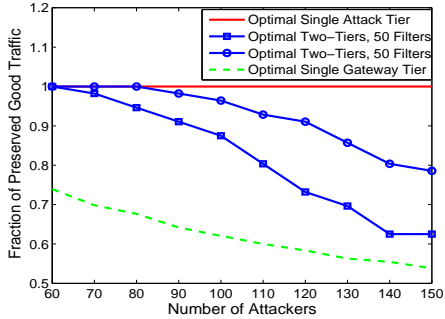

(a) % Good Traffic Preserved

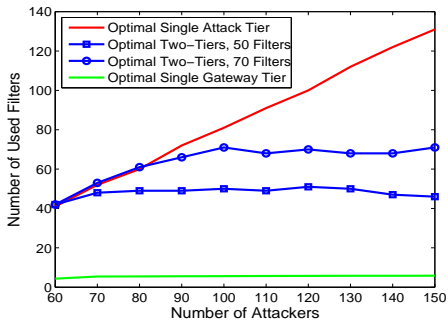

(b) Number of filters used.

**Figure 11: Performance of the Optimal Multi-tier Filtering for the Slammer scenario.**

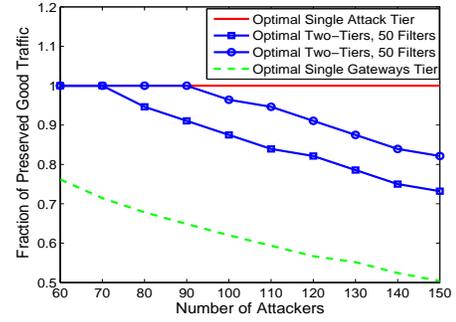

(a) % Good Traffic Preserved

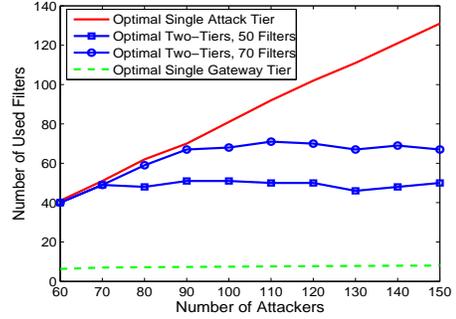

(b) Number of filters used.

**Figure 12: Performance of the Optimal Multi-tier Filtering for the Zombie scenario.**

tackers but kept the number of legitimate users fixed. The total goodput was kept constant and below the capacity, that is why it was possible to preserve up to 100% of it. We now increase the amount of good traffic together with the number of bad traffic, so it is still consistent with the Zombie scenario in Table 1.[16] In practice, such a scenario captures a combination of DDoS attack and a Flash Crowd. The results are shown in Fig.13. One can notice that while the good traffic is still below capacity, the curves are flat, meaning that we can preserve the good traffic. After the point that the good traffic exceeds the capacity, the performance degrades gracefully.

### 4.4.1 Two-Tier Heuristic (H1)

As a quick test, we tried the simple heuristic shown in Alg.3. It operates in two separate steps. In the first step, it assigns all filters optimally to the attackers-tier only. This may require $f^*_{att} > F$. In the second step, we try to correct that by filtering out (thus releasing filters) the gateways with the least amount of good traffic.

This heuristic is a very simple, two-step heuristic, which performed reasonably for the scenarios we tried. Clearly, there is much room for improvement and fine tuning, and this is the direction we are currently working on.

In this section we present some preliminary simulations of

---

[16]In particular, we increase the number of nodes at attackers tier and we assume that each node emits both good and bad traffic.

**Algorithm 3** Multi-tier Heuristic

1: Assign filters optimally at attacks' single tier. /* This may require a number of filters $f^*_{att}$ larger than the available $F^*$/
2: Order gateways in decreasing amount of good traffic $G$.
3: Allow gateways in a greedy way, i.e. from larger to smaller $G$, until you reach the maximum number of available filters $F$ (Keep the filter assignment at their attackers as assigned at the first step.).
4: Filter out all remaining gateways.

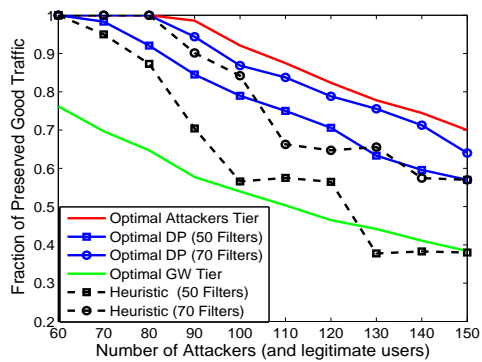

Figure 13: **Modified Zombie scenario to reflect simultaneous increase in attack traffic (DDoS) as well as in legitimate traffic (Flash Crowd).**

heuristics. The heuristic is simulated for the same scenarios we looked at for the optimal solution. In Fig.13, we presented the scenario for the combined zombie and flash crowd scenario. Together with the optimal solutions, shown in solid lines, we also show the performance of the simple heuristic in dashed lines. The heuristic is within the bounds and relatively close to the optimal for 70 filters; however its performance degrades fast with decreasing number of filters.

Given the low complexity of this simple heuristic, we are now able to simulate scenarios for a much larger number of attacks, which was prohibitively slow in simulation for the optimal solution. Fig.14 shows the results for 1000-10000 nodes for the combined DDos+flash crowd scenario. Similar trends hold as we saw before: the performance degrades and the number of required filters vary at similar rates as before.

The heuristic we explored is clearly preliminary and serve just as a proof of concept. We are currently working on improving the heuristic (i) based on the intuition we obtained from the structural properties of the optimal solution and (ii) taking into account the attack traffic distribution.

## 5. CONCLUSION

In this paper, we studied the problem of single-tier and two-tier filtering against a DDoS attack as a resource optimization problem. The purpose of filtering is to filter out individual attackers, or entire gateways, so as to maximize the amount of good traffic preserved, subject to constraints on the number of filters and the total available bandwidth. We formulated and solved the first problem as an optimization problem and showed the reduction to a well known knapsack problem. For the second problem which is a nonlinear optimization problem with non-linear constraints we showed how to solve it optimally in a dynamic programming framework and we simulated the optimal solution using realistic attack scenarios. We showed through simulations that the optimal filtering policy can bring significant improvement over any other policy in terms of preserved good traffic and number of filters used. We also developed a simple heuristic for the multi-tier scenario and showed that it performs well under realistic attack scenarios. During our simulation we have constructed models of a DDoS attack based on analysis of worm infection and showed that the optimal filtering can successfully mitigate such attack schemes. We are currently working on developing efficient heuristics to achieve near-optimal solution at lower complexity. We believe that filter optimization is both important, to enable DDoS mitigation with today's mechanisms, and an interesting problem in itself. This work serves as a quantitative upper bound of using filtering mechanisms to mitigate DDoS because it uses the optimal policy to filter attack traffic. Such a quantitative study of filtering is missing from the DDoS literature and is an essential step towards successfully mitigating DDoS attacks. One downside of filtering is that although we assumed perfect attack detection which is ideal, sometimes even optimal filtering will incur collateral damage to legitimate traffic. We are currently addressing the issue of imperfect attack identification and evaluating the performance of optimal filtering under such conditions.

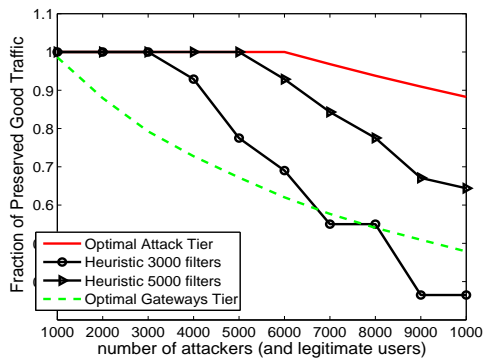

Figure 14: **Modified Zombie scenario for DDoS+Flash Crowd, for large number of nodes.**